\documentclass[]{spie}  

\usepackage{amsmath,amsfonts,amssymb}
\usepackage{graphicx}
\usepackage[colorlinks=true, allcolors=blue]{hyperref}

\usepackage{threeparttable}

\title{Characterization of sensitivity and responses of a 2-element prototype wavefront sensor for millimeter-wave adaptive optics attached to the Nobeyama 45 m telescope}

\author[a]{Satoya Nakano}
\author[a]{Yoichi Tamura}
\author[a]{Akio Taniguchi}
\author[b]{Sachiko Okumura}
\author[c]{Ryohei Kawabe}
\author[d, e]{Nozomi Okada}
\author[b]{Tomoko Nakamura}
\author[f]{Yuhei Fukasaku}

\affil[a]{Nagoya University, Nagoya, Aichi 464-8602 Japan}
\affil[b]{Japan Women's University, Bunkyo, Tokyo, 112-8681 Japan}
\affil[c]{National Astronomical Observatory of Japan, Mitaka, Tokyo, 181-8588 Japan}
\affil[d]{Osaka Metropolitan University, Sakai, Osaka, 599-8531 Japan}
\affil[e]{Ibaraki University, Mito, Ibaraki, 310-8512 Japan}
\affil[f]{University of Tsukuba, Tsukuba, Ibaraki, 305-8573 Japan}

\authorinfo{Further author information: (Send correspondence to S.N.)\\S.N.: E-mail: nasatoya@a.phys.nagoya-u.ac.jp, Telephone: +81 (0)52 789 2846}

\renewcommand{\figurename}{Figure\,}
\renewcommand{\tablename}{Table\,}
\newcommand{\sectionname}{Section\,}
\def\figref#1{\figurename\,\ref{#1}}
\def\tabref#1{\tablename\,\ref{#1}}
\def\secref#1{\sectionname\,\ref{#1}}
\def\eqref#1{(\ref{#1})}

 \usepackage{soul}

\begin{document} 
\maketitle

\begin{abstract}
We report the results of the performance characterization of a prototype wavefront sensor for millimetric adaptive optics (MAO) installed on the Nobeyama $45~\mathrm{m}$ radio telescope. MAO is a key component to realize a future large-aperture submillimeter telescope, such as Large Submillimeter Telescope (LST) or Atacama Large Aperture Submillimeter Telescope (AtLAST). The difficulty of MAO is, however, real-time sensing of wavefront deformation with $\mathrm{\sim} 10~\text{\textmu}\mathrm{m}$ accuracy across the aperture. Our wavefront sensor operating at $20~\mathrm{GHz}$ measures the radio path length between a certain position of the primary mirror surface to the focal point where a $20~\mathrm{GHz}$ coherent receiver is placed. With the 2-element prototype, we sampled two positions on the primary mirror surface (at radii of $5~\mathrm{m}$ and $16~\mathrm{m}$) at a sampling rate of $10~\mathrm{Hz}$. Then an excess path length (EPL) between the two positions was obtained by differentiating the two optical paths. A power spectral density of the EPL shows three components: a low-frequency drift ($1/f^n$), oscillations, and a white noise. A comparison of EPL measurements under a variety of wind conditions suggests that the former two are likely induced by the wind load on the telescope structure. The power of the white noise corresponds to a $1\sigma$ statistical error of $8~\text{\textmu}\mathrm{m}$ in EPL measurements. The $8~\text{\textmu}\mathrm{m}$ r.m.s.\ is significant with respect to the mirror surface accuracy required by the LST and AtLAST ($\mathrm{\sim}20\text{--}40$~$\text{\textmu m}$~r.m.s.), which demonstrates that our technique is also useful for the future large-aperture submillimeter telescopes.
\end{abstract}


\section{INTRODUCTION}
\label{sec:introductiom}

Planned large-aperture submillimeter single-dish telescopes, such as the Large Submillimeter Telescope\cite{2016SPIE.9906E..26K} (LST) and the Atacama Large Aperture Submillimeter Telescope\cite{2020SPIE11445E..2FK} (AtLAST), accommodate a wide variety of focal plane instruments and offer capabilities that allow wide field, wide instantaneous spectral coverage, and high cadence observations, which are complementary to the Atacama Large Millimeter/submillimeter Array (ALMA).
However, the difficulty in building these large-aperture telescopes is that they are largely affected by the external environment, such as wind, preventing them from achieving the expected performance.
The adaptive optics operating at millimeter wavelengths (millimetric adaptive optics\cite{2020SPIE11445E..1NT}: MAO) is one of the important components to realize such large submillimeter telescopes, while no wavefront sensing technique was available.

In Ref.~\citenum{2020SPIE11445E..1NT}, we proposed the concept of a wavefront sensor that measures the real-time variation of the radio wave path length (excess path length: EPL) from a point on the primary mirror surface to the telescope focus to realize MAO that corrects for real-time for wavefront degradation caused by wind and thermal loads on a submillimeter telescope. 
We also reported the design and lab evaluation of a prototype two-element wavefront sensor and its commissioning on the Nobeyama $45~\mathrm{m}$ radio telescope. In lab evaluation, the measurement accuracy of our wavefront sensor was estimated to be $40~\text{\textmu m}$ r.m.s.\ or better, while the performance is yet to be characterized.

In this paper, we report the results of the performance characterization, especially the noise properties and sensitivity, of the two-element wavefront sensor with the Nobeyama $45~\mathrm{m}$ radio telescope.
In \secref{sec:experiment}, we describe the overview of the experiment conducted with the Nobeyama $45~\mathrm{m}$ radio telescope in 2020.
\secref{sec:results} is devoted to the result of the experiment.
In \secref{sec:discussions}, we compare the time variation in the EPL with the wind speed and the acceleration measured by accelerometers attached to the backup structure of the primary mirror, and calculate the accuracy of the EPL measurement.
Finally, the conclusions of this paper are described in \secref{sec:conclusion}.

\begin{figure}[t]
 \centering
 \includegraphics[width=7cm,clip]{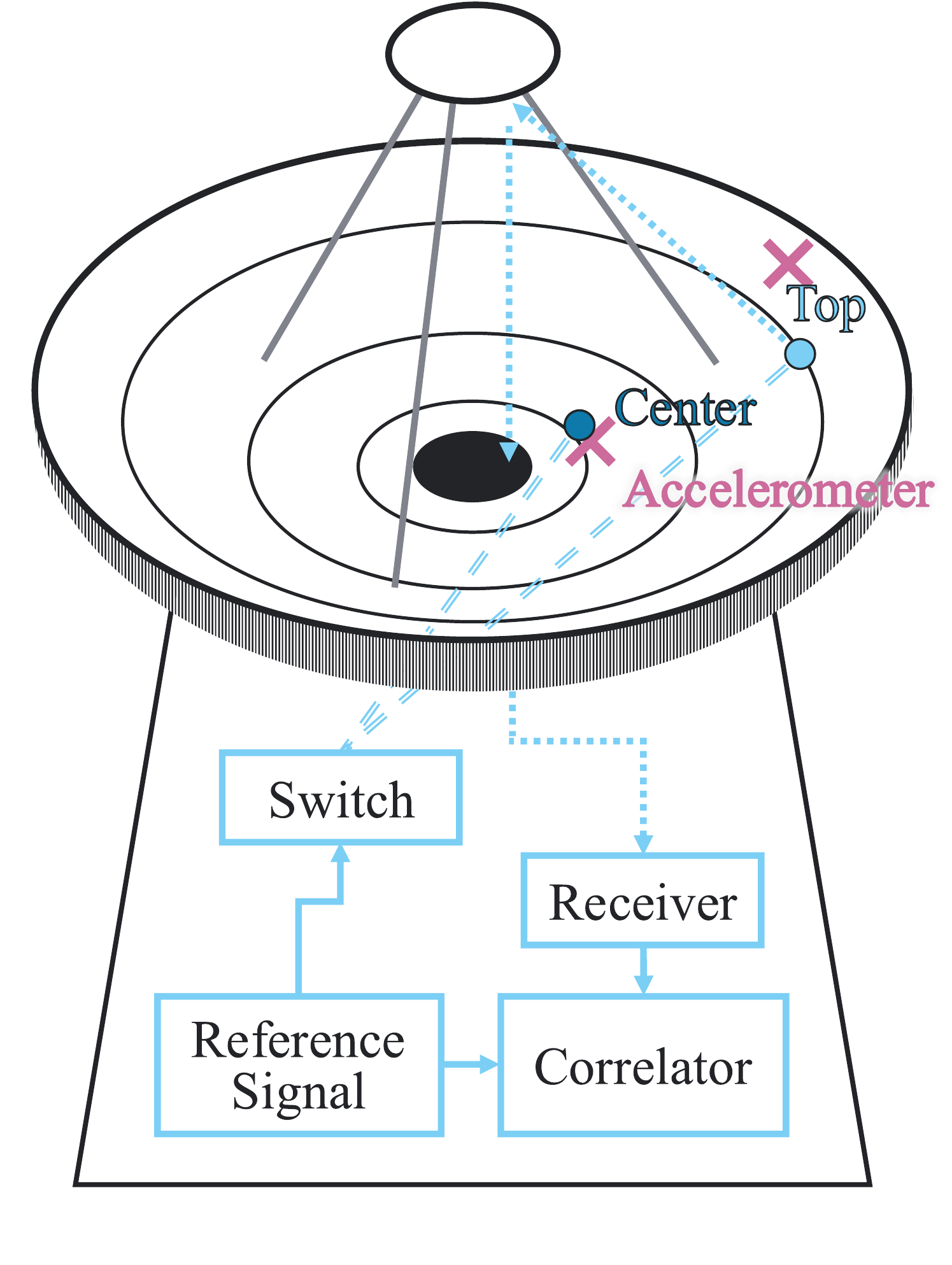}
 \caption{Schematic diagram of the experimental setup for measurements with a two-element prototype wavefront sensor with the Nobeyama 45~m radio telescope.}
\label{setup_SPIE}
\end{figure}

\section{EXPERIMENT}
\label{sec:experiment}
The concept of our wavefront sensor is aperture-plane interferometry, which is based on radio interferometry commonly used in radio astronomy. See Ref.~\citenum{2020SPIE11445E..1NT} for a detailed wavefront sensor overview and measurement principle.

\figref{setup_SPIE} represents a schematic diagram of the experimental setup.
In this experiment, two radiators were installed at two reference positions: one close to the vertex hole (``Center''; 5~m from the center of the antenna) and the other at a larger radius (``Top''; 16~m from the center of the antenna).
A broadband ($17.3\text{--}23.6~\mathrm{GHz}$) noise was used for the reference signal and was divided into two paths, one of which is sent to the radiators on the primary mirror and the other directly fed into a correlator; then, EPL was calculated from the phase difference between the two signals with the correlator.
The ``Center''-to-focus and ``Top''-to-focus EPLs were measured alternately at a switching frequency of 10~Hz. We then compute the difference between the two EPLs, $\Delta$EPL, to measure the \emph{wavefront slope} between the two reference positions.
The correlator obtained cross-power spectra by accumulating for 0.01~s, thus there are five measurement points for each integration.
The first measurement point after each switching is omitted because it contains noise due to switching, hence one EPL is the average of the four measurement points.
The error of the EPL is the standard deviation of the four measurement points.
Since the $\Delta$EPL was measured at intervals of 10~Hz, it is not sensitive to deformations of frequencies faster than 5~Hz in this measurement\footnote{The EPL between the ``Center''-to-focus and the ``Top''-to-focus are measured alternately, so they are measured at intervals of 0.1~s, respectively. Therefore, their difference, $\Delta$EPL, is obtained at a frequency of 10~Hz. Fourier transforming $\Delta$EPL yields a spectrum of $\pm$5~Hz (width 10~Hz). This spectrum is positively and negatively symmetric, which essentially means that it is sensitive to the 0--5~Hz component.}.
Measurements were conducted for 5 minutes each during moderate wind conditions on November 23, 2020, 13:01--06 (UTC+09:00; average wind speed $\mathrm{\sim}4~\mathrm{m~s^{-1}}$) and during strong/gust wind conditions on November 22, 2020, 15:21--26 (UTC+09:00; average wind speed $\mathrm{\sim}9~\mathrm{m~s^{-1}}$). 
The wind speed was measured by a meteorological instrument located at a height of about 30~m, about 75~m north of the Nobeyama $45~\mathrm{m}$ radio telescope.

For comparison, accelerometers were installed on the backside of the primary mirror (the backup structure of the primary mirror; see \figref{setup_SPIE}). The sensitivity of the accelerometers worsens at lower frequencies and the $\mathrm{\geq}$0.2~Hz component cannot be detected. 
By twice integrating the acceleration measured by the accelerometers, we obtained the displacement at the positions where the accelerometers were attached. Note that the positions of the accelerometers are not exactly the same as those of the radiators; they are placed at radii of 22~m (9~m away from ``Top'') and 4~m (1~m from ``Center''). 
Then we obtain the difference between the two displacement values, $\Delta$Displacement.

\begin{figure*}[t]
 \centering
 \includegraphics[width=17.3cm,clip]{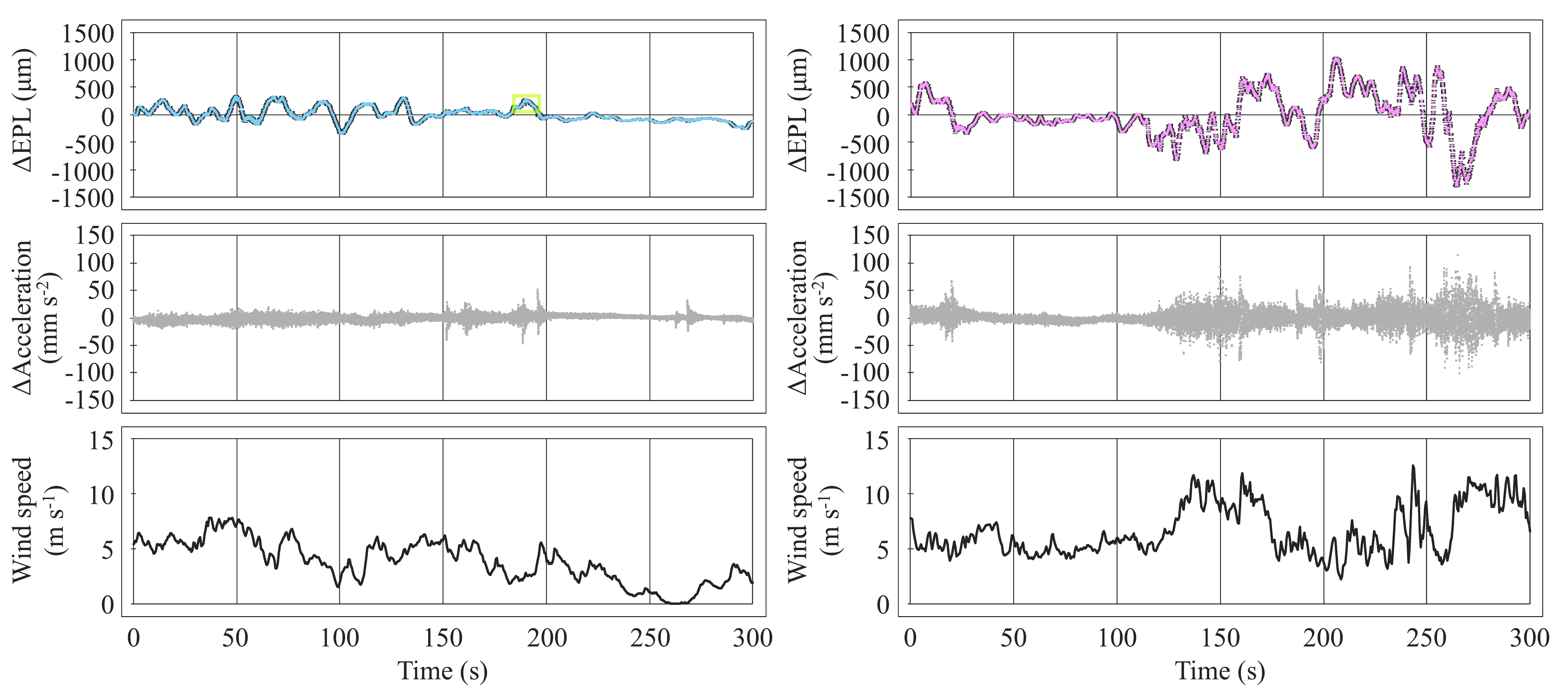}
 \caption{The results of excess path length (EPL) measurements under conditions in which the average wind speeds are moderate ($\sim$4~$\mathrm{m~s^{-1}}$, left) and high ($\sim$9~$\mathrm{m~s^{-1}}$, right). The upper panels show the excess path length at larger radius relative to the vicinity of the vertex hole plotted every 0.1~s, and the measurement error is indicated by the black line. $\Delta \mathrm{EPL}=0$ is set to the mean $\Delta$EPL averaged over the measurement time period (300~s). The green rectangle shown in the 185--195~s section of the $\Delta$EPL in weak winds represents the portion plotted in \figref{expand}. The middle panels show the acceleration at larger radius relative to the vicinity to the vertex hole plotted every 0.01~s. The lower panels show the wind speed plotted every 0.01~s.}
\label{results}
\end{figure*}

\begin{figure*}[t]
\vskip1\baselineskip
\noindent
 \centering
 \includegraphics[width=8.65cm,clip]{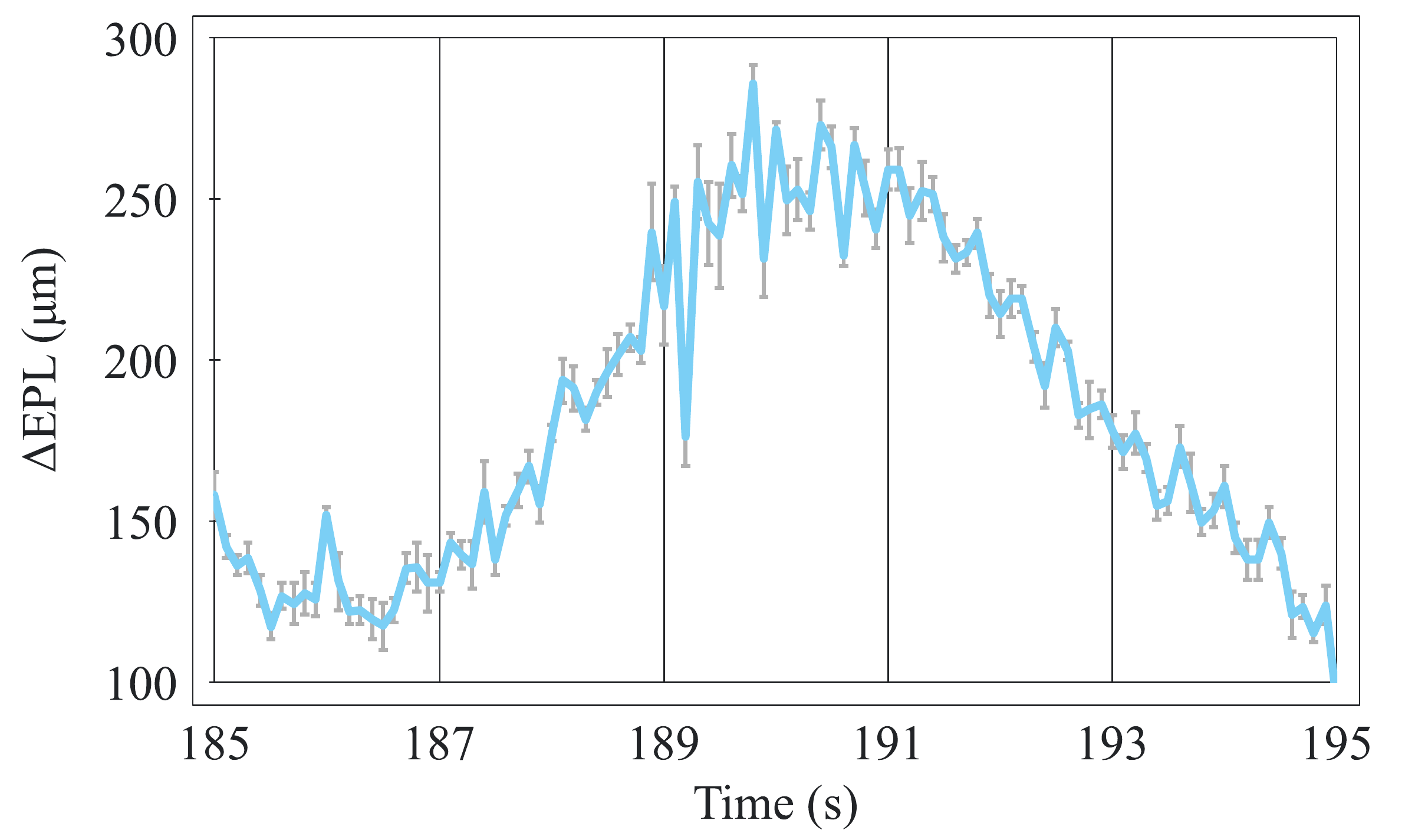}
 \caption{An enlarged 10~s plot of $\Delta$EPL under the moderate wind condition. The measurement error is indicated by the gray line. This section is indicated by the green rectangle in \figref{results}.}
\label{expand}
\end{figure*}

\section{RESULTS}
\label{sec:results}
\figref{results} shows the top-to-center EPL ($\Delta$EPL), the wind speed, and the acceleration at the top position of the primary mirror under the moderate and strong/gust wind condtions.
In addition, an enlarged plot of the measurement results for 10 seconds under the moderate wind conditions, indicated by the green rectangle in \figref{results}, is shown in \figref{expand}. Seen from the \figref{expand}, small oscillations of $\mathrm{\sim}3$~Hz period with an amplitude of several tens of \textmu m were successfully detected among the large oscillations with an amplitude of several hundreds of \textmu m. This indicates that the detection accuracy of our wavefront sensor is at least several tens of \textmu m. From \figref{results}, the amplitude of $\Delta$EPL is larger under the strong/gust wind conditions than under the moderate wind conditions, indicating that the amplitude of $\Delta$EPL tends to be larger when the wind is strong. Moreover, there seems to be a correlation between $\Delta$EPL and acceleration.
However, since our wavefront sensor and the accelerometers are capable of measuring different frequencies, it is possible that we are looking at different frequency components for acceleration and $\Delta$EPL in \figref{results}.
In \secref{sec:discussions}, we compare $\Delta$EPL and $\Delta$Displacement, which are band-pass filtered to extract common frequency components.

\section{DISCUSSIONS}
\label{sec:discussions}
\subsection{Wavefront Detection Detected by Our Wavefront Sensor}
\begin{figure*}[t]
 \centering
 \includegraphics[width=17.3cm,clip]{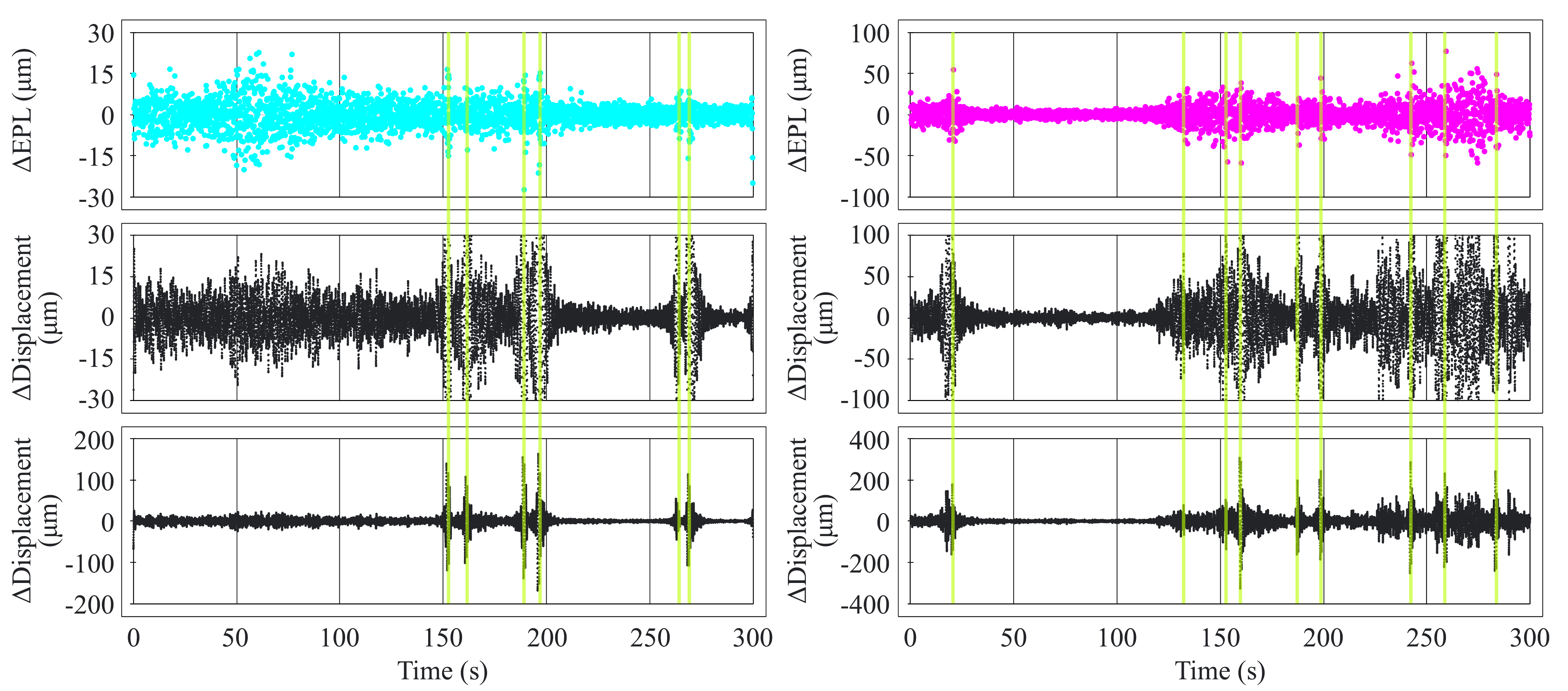}
 \caption{Comparison of $\Delta$EPL and $\Delta$Displacement under the moderate wind conditions, each bandpass filtered at 1--5~Hz, shown in cyan (left). Comparison of $\Delta$EPL and $\Delta$Displacement under the strong/gust wind conditions, each bandpass filtered at 1--5~Hz, shown in magenta (right). The upper panels show $\Delta$EPL plotted every 0.1~s. The lower panels show $\Delta$Displacement plotted every 0.01~s. The middle panels show $\Delta$Displacement scaled to $\Delta$EPL on the vertical axis. The green lines represent the remarkable peak positions, highlighting the components common to $\Delta$EPL and $\Delta$Displacement.}
\label{disp}
\end{figure*}

\begin{figure*}[t]
\vskip1\baselineskip
\noindent
 \centering
 \includegraphics[width=17.3cm,clip]{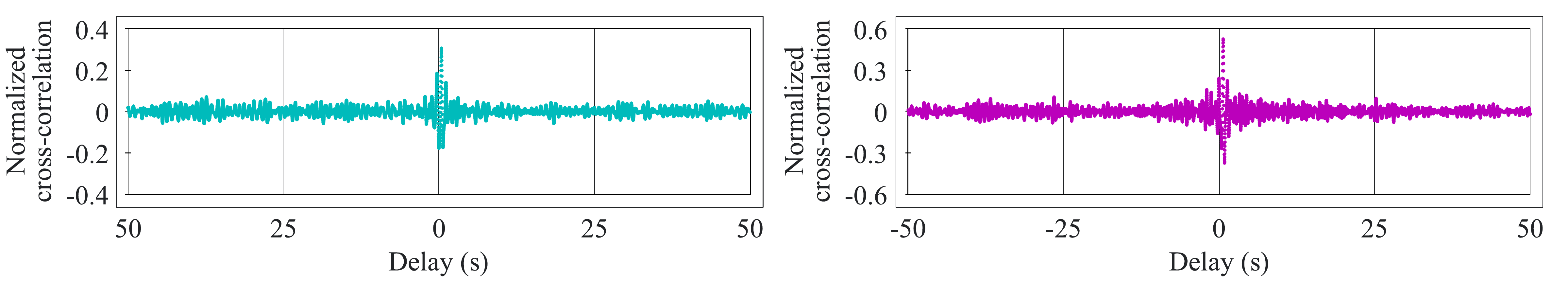}
 \caption{Normalized cross-correlation function between $\Delta$EPL and $\Delta$Displacement under the moderate wind conditions (left). Normalized cross-correlation function between $\Delta$EPL and $\Delta$Displacement under the strong/gust wind conditions (right). Since $\Delta$EPL and $\Delta$Displacement have different measurement intervals, the $\Delta$EPL measurement results is interpolated to 0.01~s intervals in this calculation. The delay times are calculated at 0.01~s intervals.}
\label{corr}
\end{figure*}

In this section, to confirm that our wavefront sensor works correctly in detecting antenna deformations, we compare the $\Delta$EPL with measurements from the accelerometers installed on the primary mirror (see \figref{setup_SPIE}).
\figref{disp} shows a comparison of $\Delta$Displacement from the accelerometers and $\Delta$EPL measured by our wavefront sensor. Since the accelerometers and our wavefront sensor are sensitive to different frequencies, we applied a band-pass filter of 1--5~Hz, respectively to extract the frequency components to which they are both sensitive.
From \figref{disp}, the envelopes seen in the scatter of $\Delta$Displacement and $\Delta$EPL are similar, and the peaks of oscillations seen in $\Delta$Displacement are also detected in $\Delta$EPL at the same time.
The normalized cross-correlation function between the band-pass filtered $\Delta$EPL and $\Delta$Displacement is shown in \figref{corr}.
From the results of \figref{corr}, the normalized cross-correlation function has a sharp peak around the delay time 0~s under the both moderate and strong/gust wind conditions.
This means that the increase and decrease of $\Delta$EPL and $\Delta$Displacement occur at the same time, suggesting that they detect the same deformation modes.
In addition, the amplitudes of $\Delta$EPL and $\Delta$Displacement shown in \figref{disp} are consistent although $\Delta$Displacement is larger under both moderate and strong/gust wind conditions.
Therefore, we conclude that $\Delta$EPLs obtained with our wavefront sensor probe wind-induced antenna deformation.

The difference found between the $\Delta$EPL and $\Delta$Displacement time-series data, such as the profiles of the envelopes and their amplitudes, can be due to the fact that they sample slightly different positions of the telescope structure. Moreover, it is possible that while $\Delta$Displacement only probes the primary mirror structure, $\Delta$EPL measures the primary mirror, the secondary, and subsequent mirrors.
In fact, the peak of the normalized cross-correlation function shows a systematic delay of 0.40~s and 0.64~s under moderate and strong/gust wind conditions, respectively, indicating that $\Delta$EPL probes actual deformation of the optics supported by the high-inertial, elastic structure of the telescope, which would exhibit the delayed, smeared deformation with respect to $\Delta$Displacement.
If this is the case, the delay could depend on what modes of deformation or oscillations are excited by the wind load.
Moreover, we also speculate that the optics support structure work as dampers since the telescope structure consists of high-inertial, elastic elements, which could only pass the low-frequency oscillations and thus reduce the pulse-like change in EPLs, as seen in \figref{disp}.
This may be one of the reasons why the amplitude of $\Delta$Displacement is larger than that of $\Delta$EPL.
Quantitative evaluation is out of scope of this paper, and we leave this for future measurements.



\begin{figure*}
 \centering
 \includegraphics[width=17.3cm,clip]{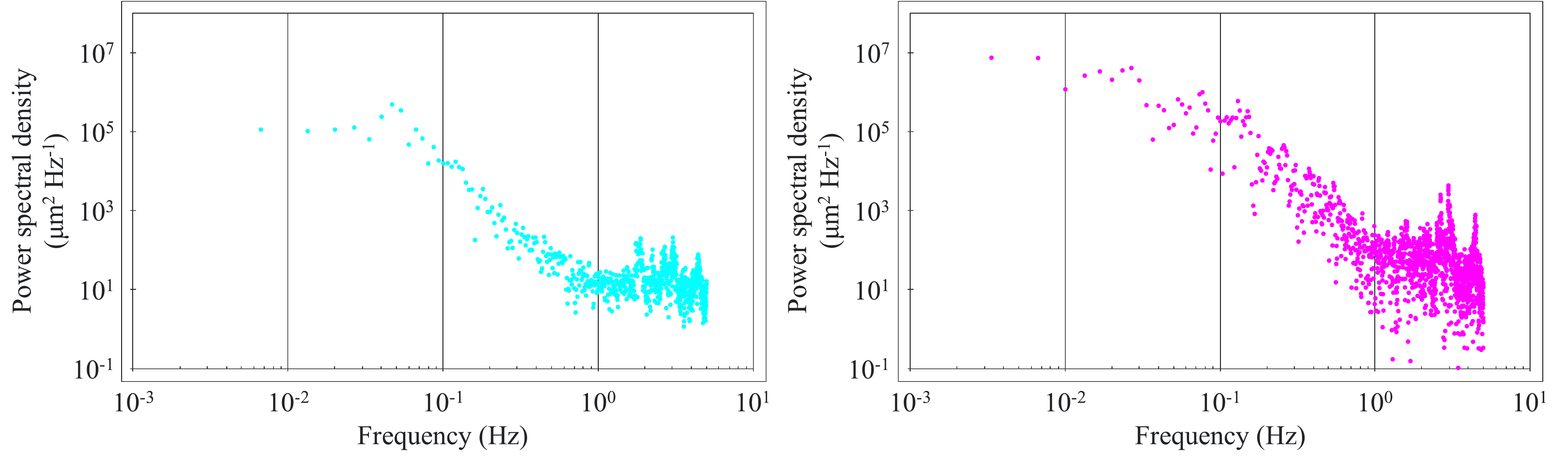}
 \caption{Power spectral density of $\Delta$EPL under the moderate wind conditions (left) and power spectral density of $\Delta$EPL under the strong/gust wind conditions (right). The vertical and horizontal axes are expressed in logarithms, respectively.}
\label{psd}
\end{figure*}

\subsection{Performance Characterization}
In this section, we analyse the $\Delta$EPL variation using the power spectral density (PSD) to characterize the output signal and noise of our wavefront sensor.
\figref{psd} shows the PSD of $\Delta$EPL obtained under the moderate and strong/gust wind conditions.
From the \figref{psd}, the $\Delta$EPL variation can be decomposed into three components: a drift component proportional to $1/f^{n}$ that dominates in the low-frequency ($<$1~Hz) region, oscillation components, and a white noise.
\figref{psd} also shows that the intensities of the $1/f^{n}$ and oscillation components increases relative to the white noise under the strong/gust wind condition, suggesting that the $1/f^{n}$ and oscillation components are excited by the wind load.

The $1/f^{n}$ component is dominant at $\mathrm{\sim}10^{-3}$--$10^{0}$~Hz and accounts for most of the total power.
Note that another systematic error can underlie in the low frequency regime, such as a long-term ($>$1~min) change in EPL due to temperature variation. Indeed, we identified a monotonous increase in EPL for a 60~min measurement of an EPL, which is consistent with the thermal expansion of the steel-made secondary mirror support in addition to the optical fiber cables expected for an ambient temperature change of $dT = \text{0.4--1.6}$~$^\circ$C (see Appendix~\ref{sec:thermal}).
For the intrinsic change in EPL, frequent delay calibration will keep the measurement accuracy, while a phase-stabilizing mechanism will be necessary for reference signal transmission in the future.

The oscillation components are found at $>$1~Hz, and the white noise is also seen at $>$1~Hz, where no oscillation components are found.
According to the Ref.~\citenum{2000SPIE.4015..467S}, the characteristic frequency of the Nobeyama 45~m radio telescope due to the wind load is $\sim$0.9~Hz. Ref.~\citenum{2020arXiv200305134H} also found several oscillation components excited by the wind load at 1.6, 2.6, 3.1, and 4.1~Hz with the accelerometers. 
Therefore, the oscillations with a peak at $\sim$1~Hz in \figref{psd} is consistent with the frequency of the oscillation due to the wind load.

Meanwhile, from the white noise, we obtain the statistical error in the $\Delta$EPL measurement.
If we adopt 3.5--4.5~Hz under the moderate wind conditions and 3.5--4.0~Hz under the strong/gust wind conditions as the frequencies at which white noise dominates in the PSD, the amplitudes of white noise in these ranges are calculated as 11.9~$\mathrm{\text{\textmu} m^{2}~Hz ^{-1}}$ under the moderate wind conditions and 23.1~$\mathrm{\text{\textmu} m^{2}~Hz^{-1}}$ under the strong/gust wind conditions.
We assume that this white noise exists as a noise floor in 0--5~Hz.
The increasing amplitude of the white noise in strong winds is likely due to the existence of harmonic oscillation componens and a component proportional to $1/f^{n}$ below the detection limit.
Using the value under the moderate wind conditions, the power of the white noise is 11.9~$\mathrm{\text{\textmu} m^{2}~Hz^{-1}}\times\mathrm{5~Hz}=\mathrm{59.4~\text{\textmu} m^{2}}$.
Thus, the statistical error in the $\Delta$EPL measurement is obtained as $\sqrt{\mathrm{59.4~\text{\textmu} m^{2}}}=7.71$~\textmu m.
Notably, however, the 3.5--4.5~Hz, which is assumed to be dominated by white noise, does not include systematic errors with long timescales of fluctuation.
Systematic errors with long timescales of fluctuation may include thermal expansion of the secondary mirror support struts and the optical fiber cables used to transmit the reference signal (Appendix~\ref{sec:thermal}). Instrument-derived systematic errors are sufficiently small to be ignored (Appendix~\ref{sec:instrument-derived}).
The detection accuracy obtained by our calculation is $\sim$3~$\text{\textmu m}$, and the derivation process is shown in Appendix~\ref{sec:calculation}.
The statistical error of $7.71$\,\,\textmu m r.m.s. obtained in this measurement is larger than the theoretical detection accuracy.
The reasons for the measured statistical error being larger than the theoretical value can include the existence of the oscillation components and the $1/f^{n}$ component below the detection limit in the white noise, as well as quantization loss in the correlator.

\section{CONCLUSION}
\label{sec:conclusion}
We presented the results of the performance characterization of the prototype two-element wavefront sensor with the Nobeyama 45~m radio telescope to realize MAO that corrects in real-time for wavefront degradation caused by deformation of a submillimeter telescope.
Our wavefront sensor alternately measured the optical path lengths from the positions at large radius and the vicinity of the vertex hole of the telescope to the focus, and the difference between the two path lengths, $\Delta$EPL, was calculated. 
The envelope of $\Delta$EPL has a shape similar to that of the displacement obtained from the accelerometers and has larger amplitude under the strong wind conditions, suggesting a successful detection of primary mirror deformation due to the wind load.
From the PSD of $\Delta$EPL, we found that the $\Delta$EPL variation is decomposed into three components: a drift component proportional to $1/f^{n}$, oscillation components, and a white noise.
From the white noise, we found that the statistical error of measurements was 7.71~\textmu m r.m.s. 
The $7.71$~\textmu m r.m.s.\ is significant with respect to the mirror surface accuracy required by the next-generation large telescopes ($\mathrm{\sim}20\text{--}40$~$\text{\textmu m}$~r.m.s.) such as the LST and AtLAST, which demonstrates that our technique is also useful for the future large-aperture submillimeter telescopes.

\appendix
\section{EVALUATION LONG-PERIOD SYSTEMATIC ERRORS BY ONE~HOUR MEASUREMENT WITHOUT SWITCHING}
\label{sec:thermal}
To evaluate the long-period systematic error, we show the results of 1~hour continuous measurements of ``Center''-to-focus and ``Top''-to-focus EPLs without switching with the two-element wavefront sensor.
The ``Center''-to-focus measurement was conducted on November 23, 2020, 9:45--10:45 (UTC+09:00, average wind speed $\sim$4~$\mathrm{m~s^{-1}}$), and the ``Top''-to-focus measurement was conducted on November 23, 2020, 11:05--12:05 (UTC+09:00, average wind speed $\sim$5~$\mathrm{m~s^{-1}}$).
Air temperature changes at a height of 30~m were measured with the same meteorological instrument that measured the wind speed.
The air temperature changes were fitted with a straight line to obtain the air temperature changes between the beginning and the end of the measurements.

\begin{figure*}
\vskip1\baselineskip
\noindent
 \centering
 \includegraphics[width=17.3cm,clip]{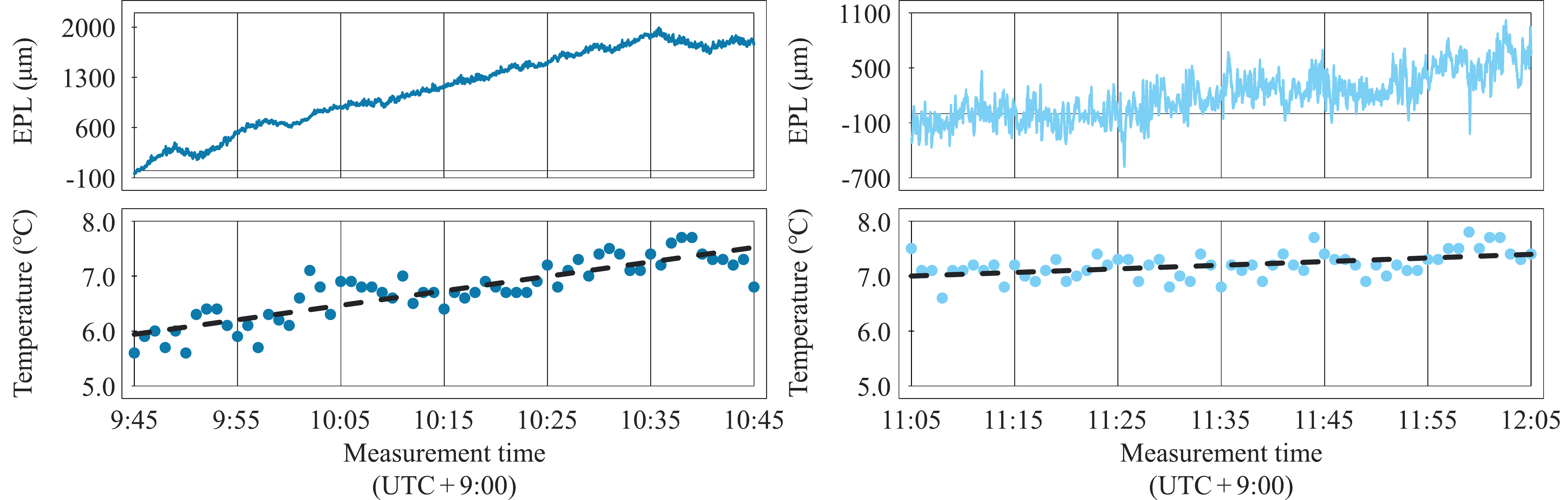}
 \caption{The results of the 1~hour ``Center''-to-focus measurement (left) and the 1~hour ``Top''-to-focus measurement. The upper panels show the excess path length plotted every 0.01~s. The average of 10--21~s is defined as $\mathrm{EPL}=0$. The lower panels show the air temperature changes at a height of 30~m plotted every 1~minute.}
\label{1h_T}
\end{figure*}

\figref{1h_T} shows the results of the 1~hour measurements of ``Center''-to-focus and ``Top''-to-focus EPLs and the air temperature changes over the measurement time.
The air temperature changes between the beginning and the end of the measurements were obtained to be $1.58\pm 0.27~\mathrm{K}$ for the ``Center''-to-focus measurement and $0.40\pm 0.21~\mathrm{K}$ for the ``Top''-to-focus measurement.
From the \figref{1h_T}, it can be seen that both ``Center''-to-focus and ``Top''-to-focus EPLs increase monotonically over 1~hour.
The increases are $\sim$1900~$\text{\textmu m}$ for ``Center''-to-focus EPL and $\sim$800~$\text{\textmu m}$ for ``Top''-to-focus EPL.
Since the measurements were conducted between morning and noon, when the temperature rises, this monotonic increases in EPLs can be attributed to thermal expansion of the telescope structure and other components.
Since it is unlikely that the primary mirror deforms on the order of millimeters in a few tens of minutes, we considered that the main cause of the monotonic increases is the thermal expansion of the secondary mirror support struts and the optical fiber cables used to transmit the reference signal.

The thermal expansion, $\Delta L$, can be evaluated by $\Delta L = \alpha L \Delta T$, where $\alpha$ is the coefficient of linear thermal expansion, $L$ is the length of the material, and $\Delta T$ is the temperature change of the material.
The approximate value of $\Delta T$ is the air temperature change shown in \figref{1h_T}.
The coefficients of linear thermal expansion and lengths of the secondary mirror support struts and the optical fiber cables are shown in \tabref{1}.
Note that since the reference signal was transmitted through the optical fiber cables, the refractive index must be considered in calculating the length.
Calculating $\Delta L$ for both the secondary mirror support struts and the optical fiber cables from the values in \tabref{1}, we obtain $\Delta L = (2.04\pm 0.35)\times 10^{3}~\text{\textmu m}$ for the ``Center''-to-focus measurement and $\Delta L = (0.52\pm 0.28)\times 10^{3}~\text{\textmu m}$ for the ``Top''-to-focus measurement.
The error is only due to the error in the air temperature changes shown in \figref{1h_T}.
Therefore, the monotonic increases of EPLs ($\sim$1900~$\text{\textmu m}$ for ``Center''-to-focus EPL and $\sim$800~$\text{\textmu m}$ for ``Top''-to-focus EPL) is consistent with the thermal expansion of the secondary mirror support struts and the optical fiber cables.

\renewcommand{\arraystretch}{1.2}
\begin{table}
\begin{center}
\begin{threeparttable}
 \centering
 \caption{Properties of the secondary mirror support struts and the optical fiber cables.}
  \label{1}
  \begin{tabular}{lcc}
    \hline
    \hline
    \shortstack{ } &
    \shortstack{Secondary mirror support struts} &
    \shortstack{Optical fiber cables} \\
    \hline
Coefficient of linear thermal expansion (K$^{-1}$) & $1.2\times 10^{-5}$ & $4.7\times 10^{-6}$ \\
Length (m) & $\sim$20 & 115 (Center), 118 (Top) \\
Refractive index & --- & 1.5 \\
\hline
  \end{tabular}
\end{threeparttable}
\end{center}
\end{table}
\renewcommand{\arraystretch}{1.0}

\section{EVALUATION OF INSTRUMENT-DERIVED SYSTEMATIC ERRORS}
\label{sec:instrument-derived}

We show the results of a 10 minutes measurement of 2~m free space in the laboratory with a one-element wavefront sensor to confirm the nature of instrument-derived systematic errors.
Instead of a telescope receiver, we used the radiator followed by a power amplifier as the receiver.
The radiator is equivalent to the one used to transmit the reference signal.
In the lab experiment, the correlator obtained cross-power spectra by accumulating for 5~ms (1/2 of the Nobeyama experiment).
To reproduce the 10~Hz switching with the two-element, the data is divided into two parts every 0.05~s.
One EPL measurement point is the average of 0.05~s measurements, and the error is the standard deviation of 0.05~s measurements. We then compute pseudo-$\Delta$EPL by taking the difference of the two divided data.
Pseudo-$\Delta$EPL corresponds to the case where two elements detect exactly the same deformation.
Thus, ideally, without instrument-derived systematic errors, pseudo-$\Delta$EPL is expected to be always zero.
Note that the instrument-derived systematic error includes fluctuations of the reference signal and correlator, but does not include systematic errors caused by different reference signal transmission paths (e.g., differences in thermal expansion of different optical fiber cables), since the one-element wavefront sensor was used.

\begin{figure*}
 \centering
 \includegraphics[width=8.65cm,clip]{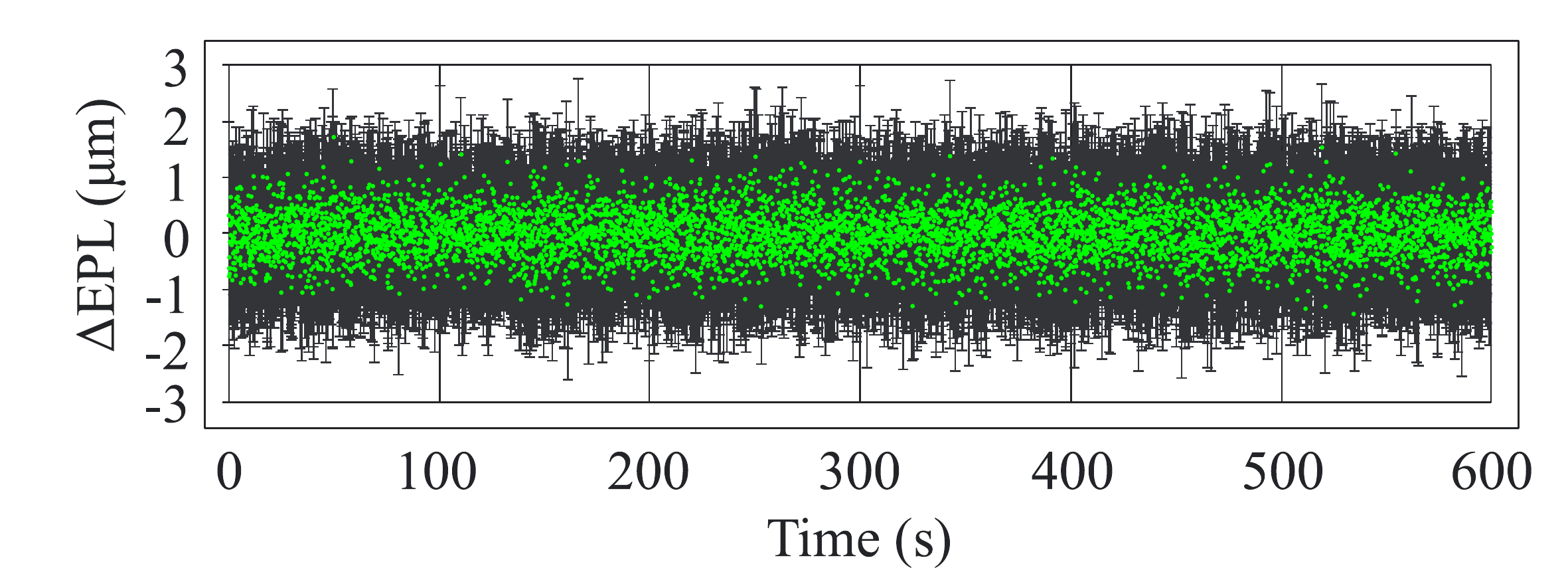}
 \caption{Pseudo-$\Delta$EPL obtained by measuring 2~m free space in the laboratory with a one-element wavefront sensor for 10 minutes, dividing the obtained data into two parts every 0.05 seconds, and taking the difference of the two divided data.
 The measurement error is indicated by the black line.}
\label{pseudo_EPL}
\end{figure*}

\begin{figure*}
\vskip1\baselineskip
\noindent
 \centering
 \includegraphics[width=8.65cm,clip]{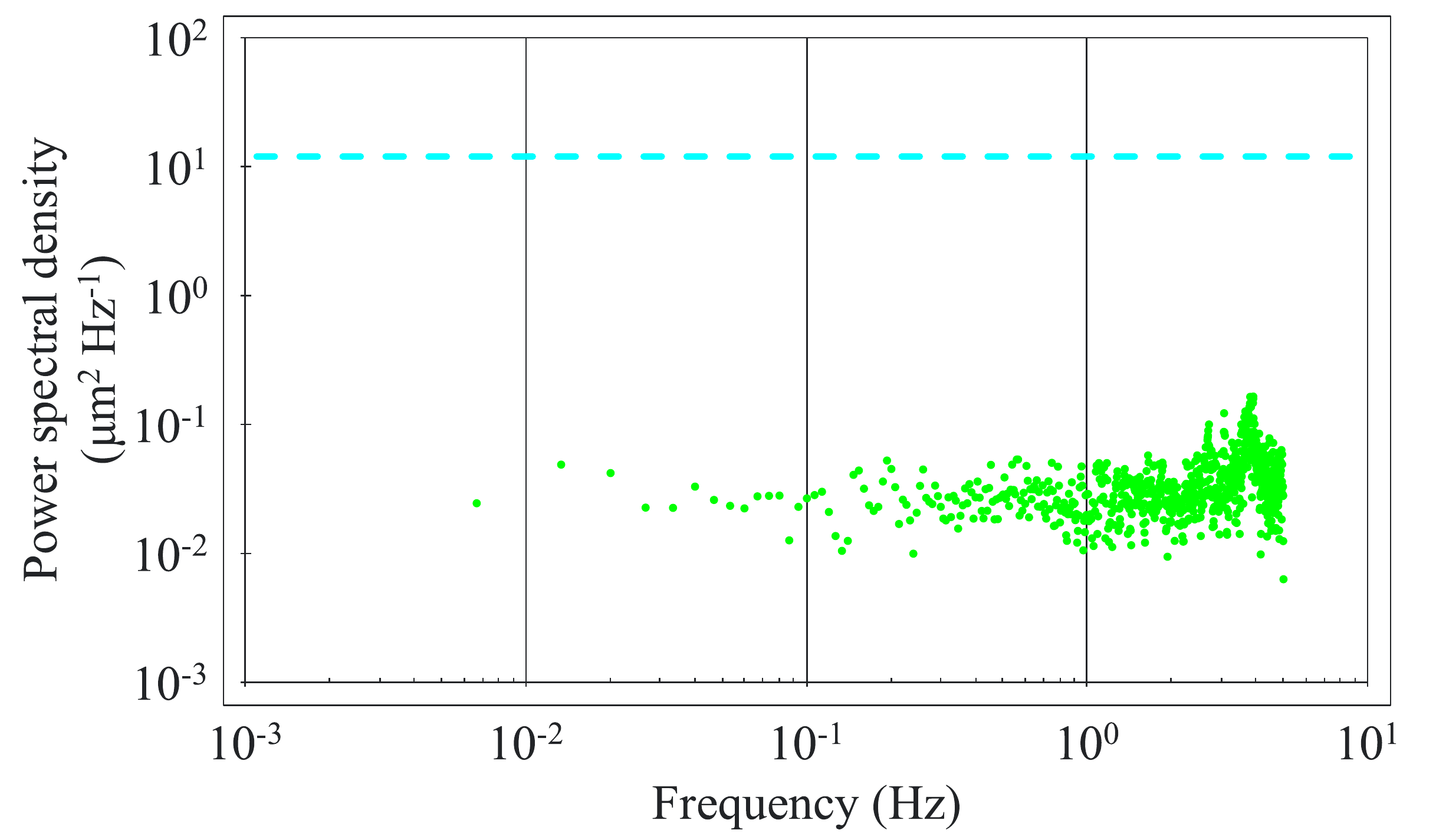}
 \caption{PSD of pseudo-$\Delta$EPL shown in \figref{pseudo_EPL}. Cyan solid line represents the amplitude of the white noise measured in the moderate wind conditions with the Nobeyama radio telescope shown in \figref{psd}.}
\label{pseudo_PSD}
\end{figure*}

\figref{pseudo_EPL} shows the result of the measurement of pseudo-$\Delta$EPL.
From \figref{pseudo_EPL}, we found that pseudo~$\Delta$EPL has much smaller amplitudes than those of $\Delta$EPL measured with the Nobeyama 45 m radio telescope shown in \figref{results}.
\figref{pseudo_PSD} shows PSD of pseudo~$\Delta$EPL shown in \figref{pseudo_EPL}.
From \figref{pseudo_PSD}, we found that PSD is constant at low frequencies ($<$2~Hz).
The slope of the $<$2~Hz section is calculated to be 0.005$\pm$0.022, which is flat within the error range.
Therefore, the instrument-derived systematic error does not include components increasing at low frequencies.
The amplitude of the white noise calculated for $<$2~Hz excluding 0~Hz is 0.0275~$\mathrm{\text{\textmu} m^{2}\,\,Hz^{-1}}$.
From this white noise, the instrument-derived systematic error is calculated to be 0.371 $\text{\textmu m}$ r.m.s.
We conclude that the instrument-derived systematic error is sufficiently smaller than the statistical error measured with the Nobeyama radio telescope ($7.71$\,\,\textmu m r.m.s.) and has little effect on the measurement.

\section{CALCULATION OF $\Delta$EPL ACCURACY}
\label{sec:calculation}
We calculate the theoretical detection accuracy of $\Delta$EPL.
As shown in the Appendix of Ref.~\citenum{2020SPIE11445E..1NT}, the sensitivity of our wavefront sensor can be expressed as 
\[
\sigma_{\theta} \approx 0.078 \ \mathrm{(deg)} \times \left (\frac{\Delta \nu} {\rm 8192~MHz} \right) ^{-0.5} \left (\frac {\Delta t} {\rm 10~ms} \right)^{-0.5} \left (\frac{N_2 / S_2}{300} \right)^{0.5},
\]
where $\Delta \nu$ is the receiver bandwidth, $\Delta t$ is the integration time, $N_{2}$ is the system noise temperature, and $S_{2}$ is the amplitude of the received reference signal.
In this measurement, we have $\Delta \nu \sim 2500~\mathrm{MHz}$, $\Delta t=40~\mathrm{ms}$, $N_{2}\sim 200~\mathrm{K}$.
The calculation of $S_{2}$ requires several steps.
First, the power measured just before the radiator was $-28.2$~dBm.
The power $P$ and antenna temperature $T_\mathrm{a}$ can be converted by $T_\mathrm{a}=P/k_\mathrm{B}\Delta B$, where $k_{\mathrm{B}}$ is Boltzmann's constant and $\Delta B$ is the signal bandwidth. 
Since the reference signal is a broadband noise of $17.3\text{--}23.6\,\,\mathrm{GHz}$, we have $\Delta B\sim 6~\mathrm{GHz}$.
Therefore, the amplitude of the signal before the radiator in units of K is $T=1.74\times 10^{7}~\mathrm{K}$.
Meanwhile, using the chopper wheel calibration with the Nobeyama 45~m radio telescope, we obtained the optical coupling factor between the radiator and the receiver to be 71.28~dB.
Hence, the amplitude of the received signal in units of K is calculated to be $S_{2}\sim1.3~\mathrm{K}$.
Substituting these values into the equation, we obtain $\sigma _{\theta} \sim 5.6\times 10^{-2}~\mathrm{deg}$.
Using the speed of light $c$ and the frequency $\nu=20~\mathrm{GHz}$, the phase difference and EPL have the relationship $\Delta \phi =2\pi \frac{\mathrm{EPL}}{c}\nu$.
Thus, $5.1\times 10^{-2}~\mathrm{deg}$ corresponds to a detection accuracy of $2.1~\text{\textmu m}$ in EPL.
Since the $\Delta$EPL is calculated by taking the difference of two EPLs, its detection accuracy is expected to be $\sim$3~$\text{\textmu m}$, considering error propagation.
Notably, due to the roughly estimated receiver noise amplitude $N_{2}$ and receiver bandwidth $\Delta \nu$, the calculated detection accuracy is only as reliable as one significant digit.

\acknowledgments 

S.N.\ is supported by Iwadare Scholarship Foundation.  
This study is supported by JSPS KAKENHI Grant (No.~17H06206 and 17H06130) and NAOJ Research Coordination Committee, NINS (NAOJ-RCC-2201-0102). This study was also supported by a grant from the Hayakawa Satio Fund awarded by the Astronomical Society of Japan.

\bibliography{report} 
\bibliographystyle{spiebib} 

\end{document}